\documentclass[reprint, preprintnumbers, aps, prd, floatfix, superscriptaddress, nofootinbib]{revtex4-2}

\usepackage{silence}
\usepackage{amssymb, amsmath, graphicx, xcolor, hyperref, orcidlink, fontawesome}

\newcommand\Msun{\ensuremath{M_{\odot}}}

\newcommand{\IoA}{Institute of Astronomy, University of Cambridge, Madingley Road, Cambridge, CB3 0HA, UK}
\newcommand{\KICC}{Kavli Institute for Cosmology, University of Cambridge, Madingley Road, Cambridge, CB3 0HA, UK}
\newcommand{\DAMTP}{Department of Applied Mathematics and Theoretical Physics, Centre for Mathematical Sciences, University of Cambridge, Wilberforce Road, Cambridge, CB3 0WA, UK}

\begin{document}

\title{Modeling the LISA response in the time-frequency domain}

\author{Tomasz Kinowski\,\orcidlink{0009-0004-9474-7654}}
\email{tk593@ast.cam.ac.uk}
\affiliation{\IoA} \affiliation{\KICC}

\author{Christopher J.\ Moore\,\orcidlink{0000-0002-2527-0213}}
\email{cjm96@cam.ac.uk}
\affiliation{\IoA} \affiliation{\DAMTP} \affiliation{\KICC}

\date{\today}

\begin{abstract}
    Time-frequency methods are a promising approach to LISA data analysis. 
    This paper presents a model for the LISA response directly in the time-frequency domain using the Wilson-Daubechies-Meyer (WDM) transformation. 
    The implementation is vectorized using \texttt{JAX} and can efficiently evaluate the response for batches of signals, achieving white-noise mismatches of $\sim \! 10^{-9}$ compared to existing implementations of the LISA one-arm response.
    Our results demonstrate that the LISA response can be modeled accurately and efficiently in the time-frequency domain, providing an important component of an end-to-end time-frequency LISA analysis framework.
\end{abstract}

\maketitle

\section{Introduction} \label{sec:intro}

The Laser Interferometer Space Antenna (LISA) will be one of the first space-based gravitational-wave (GW) detectors expected to launch in the mid-2030s \cite{2024arXiv240207571C, 2017arXiv170200786A, 2016CQGra..33c5010L, 2020IJMPA..3550075R}. 
It will consist of three spacecraft flying in a triangular configuration separated by $\ell \! \approx \! 2.5\times 10^6\,\mathrm{km}$ on an $R \! = \! 1\,\mathrm{au}$ heliocentric orbit. 
With sensitivity to millihertz frequencies, LISA will observe new GW sources \cite{2023LRR....26....2A}, many of which will produce long-duration signals with complex frequency structure, necessitating the development of new data-analysis techniques. 

The signals observed by the ground-based LIGO and Virgo detectors are short compared with the timescales on which the detectors move; the detector response can therefore be treated as time-independent. 
Current detectors also operate in the long-wavelength limit---the detector arm length is small compared with the GW wavelength---so that the response is frequency-independent \cite{1973grav.book.....M, 1987thyg.book..330T,2008CQGra..25r4017R}. 
Consequently, the detector response can be modeled in either the time or frequency domain simply by multiplying by an antenna pattern function.

Additionally, over short durations, instrumental noise can be approximated as stationary (e.g.\ Refs.~\cite{2016CQGra..33m4001A,2020CQGra..37e5002A}). 
Consequently, the noise covariance is approximately diagonal in the frequency domain. 
This combination of factors makes the frequency domain the natural representation for analyzing data from ground-based detectors.

This contrasts strongly with the situation for LISA.
Modeling the LISA response requires accounting for the effect of GWs on each of the six one-arm laser links as well as the synthesis of the time-delay interferometry (TDI) variables used to suppress laser frequency noise \cite{PhysRevD.59.102003,1999ApJ...527..814A,PhysRevD.62.042002,2004PhRvD..70j2003N,maggiore,2021LRR....24....1T}. 
Many LISA sources have multi-year durations during which the spacecraft move appreciably, making the response time dependent. 
Additionally, the long-wavelength approximation breaks down at the high-frequency end of the LISA band \cite{2024arXiv240207571C,2003PhRvD..67b2001C,2004PhRvD..70h2002V,2004PhRvD..69h2003R}; the transfer functions describing the one-arm responses are frequency dependent. 
Existing methods for computing the LISA response account for this while continuing to operate in either the time or frequency domain \cite{2022PhRvD.106j3001K, bayle_2025_17238548, 2020PhRvD.102b3033K, 2022PhRvD.105d4055K, katz_2021_5730688}. 
However, the additional challenge of modeling non-stationary noise sources motivates investigating time-frequency (TF) methods for LISA data analysis \cite{2020PhRvD.102l4038C}. 

In the frequency domain, sine waves provide the natural basis for the Fourier transform. 
However, there is no preferred basis in the TF domain.
Many wavelet methods generate TF bases through translations and dilations of a single ``mother wavelet'', leading to a dyadic tiling of the TF plane in which the time resolution increases with frequency at the expense of frequency resolution. 
These methods have found many applications across signal processing, quantum physics, machine learning, and other fields \cite{Guo2022WaveletReview,1999RvMP...71..267A}.
In the context of LISA, Ref.~\cite{2020PhRvD.102l4038C} identified a Wilson-Daubechies transformation with a Meyer window function (hereafter, simply the Wilson-Daubechies-Meyer (WDM) transformation) as generating a TF basis particularly suitable for GW data analysis. 
(See also Refs.~\cite{2026arXiv260621473J, 2026arXiv260620269V, 2026PhRvD.114d4035C, 2026arXiv261019002J,2026arXiv261117396M}.) 
The WDM transformation provides a uniform tiling of the TF plane (see Fig.~\ref{fig:gnm_examples}), in which the basis wavepackets have fixed resolution in time and frequency satisfying $\Delta T \Delta F \! = \! 1/2$. 
The WDM basis has a long history in GW data analysis, having been introduced by Ref.~\cite{2012JPhCS.363a2032N} and used by coherent WaveBurst (cWB) for analyzing unmodeled transients \cite{2021SoftX..1400678D}. 
Other TF bases can also be used; for example, Ref.~\cite{2026arXiv260905601P} considered a Wilson-Daubechies-Gaussian (WDG) basis with slightly different localization in time and frequency.

\begin{figure*}[t]
    \centering
    \includegraphics[width=0.95\textwidth]{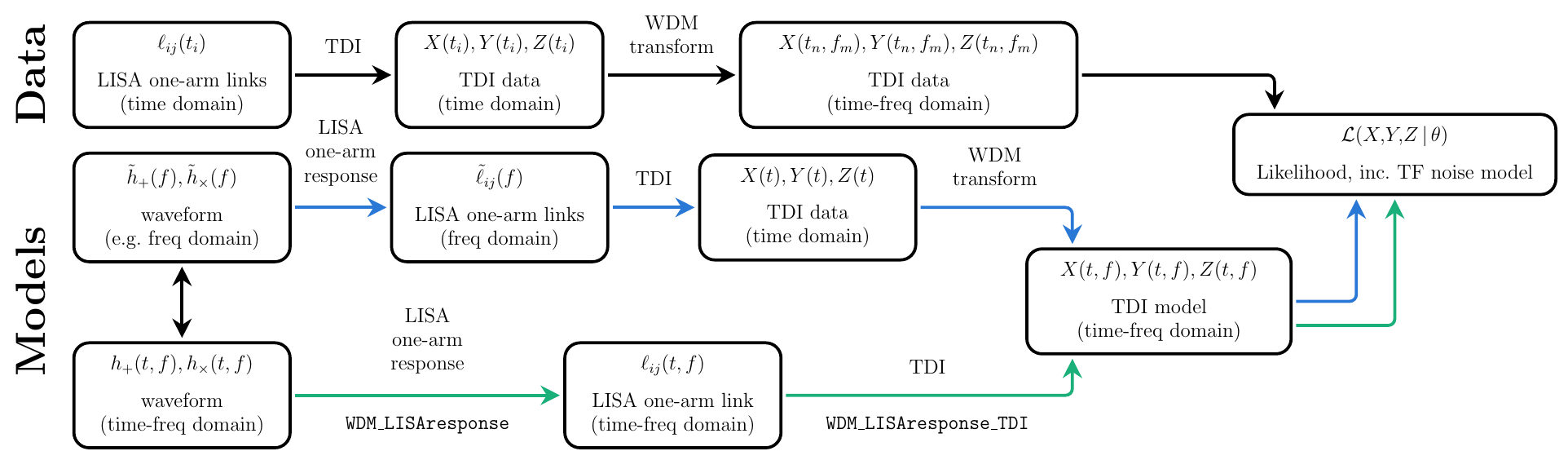}
    \caption{ \label{fig:flow_chart} 
        A flow chart of the steps to evaluate the LISA likelihood in the TF domain.
        The likelihood (final box) compares the data with models, both in the TF domain.
        \textbf{The data:} one-arm link measurements are combined into TDI variables before being transformed into the TF domain with a discrete WDM transform (black pathway).
        \textbf{The models:} these can be handled in several ways.
        Given a frequency-domain waveform, one approach is to evaluate the LISA response before transforming into the TF domain (blue pathway).
        Alternatively, if the waveform is available in the TF domain, calculating the LISA response in the TF domain avoids the need for a discrete WDM transform (green pathway); this is the subject of this paper.
    }
\end{figure*}

There are several advantages to a TF representation.
Noise that varies slowly in both time and frequency can be represented with an approximately diagonal covariance matrix in the TF domain, allowing the stationary-noise assumption to be relaxed more easily than in the frequency domain \cite{2020PhRvD.102l4038C, 2026PhRvD.114d4035C, 2026arXiv261019002J}. 
This can speed up the calculation of the likelihood (see the final box in Fig.~\ref{fig:flow_chart}).
Short-duration gaps and glitches can also be handled by masking out the corresponding columns in the TF grid (taking care to account for leakage between adjacent time bins \cite{2012JPhCS.363a2032N,2020PhRvD.102l4038C,2026PhRvD.113f4033P}). 
Additionally, the TF representation naturally motivates several approximations that can be used to accelerate the response model.
Finally, efficient methods exist for generating a range of waveforms in the TF domain \cite{2020PhRvD.102l4038C,2026arXiv260928316C} (the double-headed arrow in Fig.~\ref{fig:flow_chart}).

For the LISA likelihood, one approach is to compute one-arm response and TDI variables in the time or frequency domain and then apply a discrete WDM transform to express them in the TF domain;
this is represented by the blue pathway in Fig.~\ref{fig:flow_chart}.
However, this differs from the standard structure of GW data analysis, where the likelihood is evaluated end-to-end in a single representation (usually the frequency domain) without the need for multiple fast Fourier transforms (see, for example, Refs.~\cite{2019ApJS..241...27A,2019PASA...36...10T,2022RvMP...94b5001C}). 
It is desirable to be able to do the same for LISA and to model the one-arm response and TDI variables directly in the TF domain. 
This may also allow us to exploit sparse representations of these quantities in the TF domain. 
Motivated by these considerations, we seek an end-to-end pipeline for the LISA likelihood in the TF domain; that is, we seek to follow the green pathway in Fig.~\ref{fig:flow_chart}. 

This paper presents a model for the LISA response in the TF domain that is made publicly available in the \texttt{WDM\_LISAresponse} package  \href{https://github.com/tomaszkinowski/WDM_LISAresponse}{\faGithub}. 
Sec.~\ref{sec:summary} contains a high-level overview of our approach and some key results.
Sec.~\ref{sec:TF_data_analysis} summarizes key details of the WDM transform as implemented in Ref.~\cite{2026arXiv261117396M}. 
Sec.~\ref{sec:response} derives the TF-domain LISA response. 
Sec.~\ref{sec:results} compares our response model with existing methods. 
Finally, Sec.~\ref{sec:discussion} outlines the work that remains to develop a workable, end-to-end TF-domain LISA data-analysis pipeline.

We use units in which $c \! = \! 1$ throughout, so the LISA arm length is $\ell \! \approx \! 8\,\mathrm{s}$ and the orbital radius is $R \! \approx \! 500\,\mathrm{s}$.

\section{Executive Summary} \label{sec:summary}

The response of a detector maps the incoming GW polarizations $h_{+}$ and $h_{\times}$ to the measured detector output.
For LIGO and Virgo, the measured output is the differential arm length and the response can be simply modeled by multiplying the GW polarizations by time- and frequency-independent antenna pattern functions.

\begin{figure}[b]
    \centering
    \includegraphics[width=0.46\textwidth]{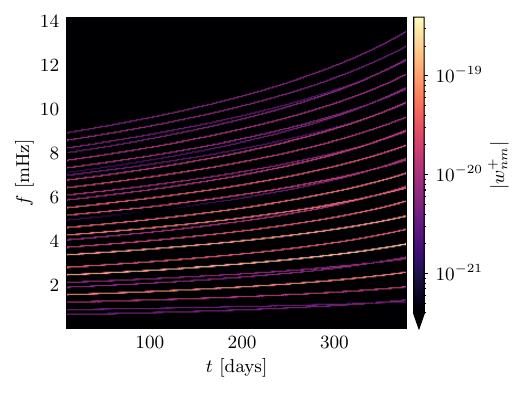}
    \caption{ \label{fig:fig-emri-tracks} 
        WDM transform of $h_+$ for an EMRI signal with the parameters in Tab.~\ref{tab:few_parameters}. 
        The harmonic tracks are clearly visible.
    }
\end{figure}

For LISA, the measured output is usually taken to be the second-generation TDI channels $X_2$, $Y_2$, and $Z_2$, and the response is more complicated because it depends on both time and frequency.
Our TF approach uses the WDM transformation to decompose the waveform polarizations into a sum of wavepackets.
These wavepackets are localized in time and frequency, with typical duration $\Delta T$ and bandwidth $\Delta F \! = \! (2\Delta T)^{-1}$.
These pixels, or atoms, uniformly tile the TF plane and provide a convenient, sparse representation for many GW signals.
An example WDM transformation of an extreme-mass-ratio inspiral (EMRI) signal is shown in Fig.~\ref{fig:fig-emri-tracks}; the multiple harmonic tracks at multiples of the orbital frequencies are clearly visible.
When taking a WDM transformation, we are free to choose a suitable time resolution, $\Delta T$.

We model the LISA response to individual wavepackets.
Provided $\Delta T \! \ll \! 1\,\mathrm{year}$, the timescale of the LISA orbit, we can approximate the detector as stationary over the wavepacket duration; we refer to this as the \emph{short-atom approximation}.
Similarly, provided $\Delta F \! \ll \! 1/\ell$, the inverse light-travel time between the LISA spacecraft, we can approximate the transfer function as frequency-independent across the wavepacket bandwidth; we refer to this as the \emph{narrowband approximation}.
At zeroth order, these approximations remove the time and frequency dependence, respectively, from the LISA response.

One can picture this as approximating LISA as a simple LIGO-like detector for each wavepacket.
The response can then be modeled by multiplying the GW polarizations by suitable antenna pattern functions, which depend on the central time and frequency of each wavepacket.
Our implementation vectorizes these operations using \texttt{JAX}, allowing the response to be evaluated efficiently for batches of signals on GPUs.

\begin{figure}[t]
    \centering
    \includegraphics[width=0.49\textwidth]{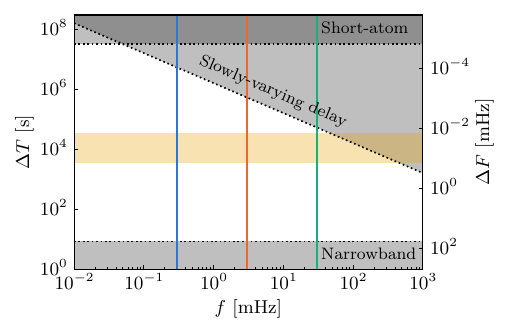}
    \caption{ \label{fig:valid_range}
        The choice of WDM resolution, $\Delta T$, is constrained by the three approximations described in the text.
        The \emph{short-atom approximation} requires $\Delta T \! \ll \! 1\,\mathrm{year}$; below the upper line (shaded regions are prohibited).
        The \emph{narrowband approximation} requires $\Delta T \! \gg \! \ell$; above the lower line. 
        The \emph{slowly-varying delay approximation} requires the frequency-dependent condition $\Delta T \! \ll \! (2\pi f_{\rm GW}|\dot{\delta}|)^{-1}$; the sloped line. 
        These are not hard boundaries,  rather the error grows as the dotted line is approached from the valid region.
        The yellow band indicates resolutions for which our response is expected to be accurate across the LISA band while giving a suitable frequency resolution.
        Vertical lines indicate frequencies used to test the response throughout this paper, see Sec.~\ref{sec:results}.
    }
\end{figure}

\begin{figure}[t]
    \centering
    \includegraphics[width=0.49\textwidth]{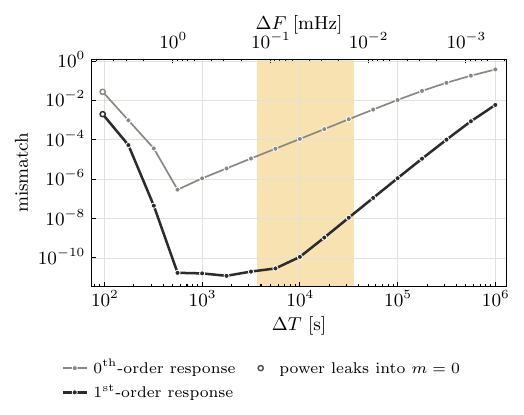}
    \caption{ \label{fig:fig-improved-accuracy} 
        The white-noise mismatch in the one-arm LISA response for the example EMRI signal plotted as a function of the WDM resolution $\Delta T$.
        The mismatch compares the TF response with a reference time-domain model.
        The yellow band shows the suitable range of $\Delta T$ from Fig.~\ref{fig:valid_range}.
        The two lines show the mismatches obtained by the zeroth- and first-order response models described in the text; the first-order model is highly accurate throughout the recommended range of $\Delta T$.
        The first-order EMRI results also appear in Sec.~\ref{sec:results}, where additional information is given.
    }
\end{figure}

Finally, the responses of the individual wavepackets must be recombined.
This requires time shifting the wavepackets by a delay $\delta$ of characteristic size $\sim \! R$ to realign them. 
This is necessary because the effective LIGO-like responses associated with each wavepacket are centered at different points on LISA's orbit.
This realignment is a key driver of the overall computational cost of the response.
Time shifts of WDM wavepackets are evaluated using the time-delay filter functions described in Sec.~\ref{subsec:td_filters}.
The delay can be approximated as constant if the detector does not move appreciably within the duration of a wavepacket; this requires expanding in the small quantity $\Delta T \! \ll \! (2\pi f_{\rm GW}|\dot{\delta}|)^{-1}$ which is referred to as the \emph{slowly-varying delay approximation}.

The validity of these three approximations constrains the choice of $\Delta T$.
Since $\Delta F \! = \! 1/(2\Delta T)$, the frequency resolution cannot be varied independently.
Nevertheless, for typical LISA sources there is a broad range of $\Delta T$ over which all three approximations are simultaneously valid.
These ranges of validity are illustrated in Fig.~\ref{fig:valid_range}.
By choosing $\Delta T\approx 10^{4}\,\mathrm{s}$ (in the highlighted band) it is possible to satisfy all three approximations for GW frequencies across the LISA bandwidth. 
Conveniently, this choice also gives a sub-millihertz frequency resolution, $\Delta F \! = \! (2\Delta T)^{-1} \! \approx \! 5\times 10^{-2}\,\mathrm{mHz}$, which is suitable for modeling the smoothly varying part of the LISA noise power-spectral density.
The fact that LISA's noise properties are expected to drift over timescales of a few days (and the confusion noise over a few months) also makes $\Delta T \! \approx \! 10^4\,\mathrm{s}$ a suitable resolution for modeling the noise.

A naive implementation of the response with the above approximations performs moderately well; hereafter, we will refer to this as the zeroth-order response model.
Mismatches are as low as $\mathcal{M} \! \approx \! 10^{-5}$ at the low-frequency end of the LISA band, gradually degrading to $\mathcal{M} \! \approx \! 10^{-2}$ at the high-frequency end; see Fig.~\ref{fig:tiling-sweep-zeroth} in Appendix~\ref{sec:accuracy_of_zeroth_order_response}.

However, several further improvements are possible.
Each of the aforementioned approximations can be carried out to higher order.
For example, rather than treating LISA as static over the duration of each wavepacket in the \emph{short-atom approximation}, we can include its linear motion, leaving corrections of $\mathcal{O}((\Delta T/\mathrm{year})^2)$.
Similarly, the \emph{narrowband} and \emph{slowly-varying delay} approximations can be carried out to first order, rather than zeroth order, in the small quantities $\ell/\Delta T$ and $\Delta T f_{\rm GW} \dot{\delta}$, respectively.

Finally, the accuracy of the wavepacket realignment can be greatly improved by evaluating the waveform at the LISA constellation barycenter rather than the solar system barycenter.
If this shifted waveform is available, the maximum required time shift is dramatically reduced from $\delta \! \approx \! R$ to $\approx \! \ell$ (from $\approx \! 500\,\mathrm{s}$ to $\approx \! 8\,\mathrm{s}$) which can be efficiently carried out to higher accuracy.
Hereafter, we will refer to this as the first-order response model; this is the version of the response that should be used.

With these improvements, the response performs exceptionally well with mismatches around $\mathcal{M} \! \approx \! 10^{-9}$ across the LISA band.
This mismatch is plotted as a function of the resolution $\Delta T$ in Fig.~\ref{fig:fig-improved-accuracy} for the example EMRI signal in Fig.~\ref{fig:fig-emri-tracks}.
These first-order six-link responses were calculated for one-year long signals on an Intel Core Ultra 7 265K CPU with 20 logical cores, in $\sim \! 360\,\mathrm{ms}$.

\section{Time-Frequency Data Analysis} \label{sec:TF_data_analysis}

\subsection{The WDM basis} \label{subsec:WDMtransform}

This section gives a brief, high-level overview of the WDM functions used as a TF basis in this paper, with the aim of building intuition and establishing notation.
We do not attempt to provide a rigorous or complete description here; for full details, we refer the reader to Ref.~\cite{2026arXiv261117396M} and Ref.~\cite{WDM_github} (see also Refs.~\cite{2012JPhCS.363a2032N, 2020PhRvD.102l4038C, 2026arXiv260621473J}).

\begin{figure}[t]
    \centering
    \includegraphics[width=0.49\textwidth]{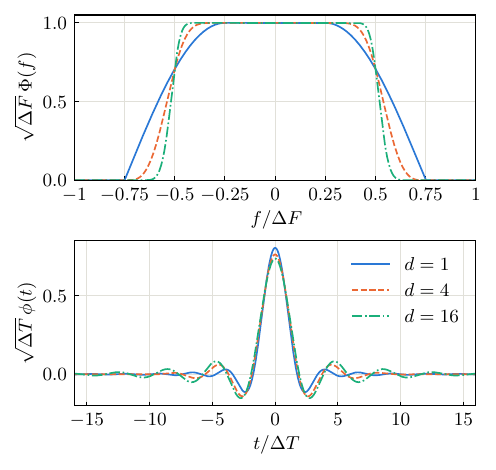}
    \caption{ \label{fig:meyer-window} 
        Dimensionless Meyer window function in the frequency (top panel) and time (bottom panel) domains for different values of $d$.
        In the frequency domain, the window has a flat-top region $[-0.25,0.25]$ corresponding to $A\!=\!0.25$.
        The window has compact support in frequency and is localized in time.
        This is based on similar figures in Refs.~\cite{2026arXiv261117396M, 2020PhRvD.102l4038C}.
    }
\end{figure}

A real-valued time series $x(t)$ is uniformly sampled at times ${t_a \! = \! a\delta t}$, where ${a \! = \! 0, 1, \ldots, N-1}$ and $T \! = \! N\delta t$ is the duration. 
The discretely sampled time series is
\begin{align} \label{eq:discrete_time_series}
    x_a \equiv x(t_a) .
\end{align}

The frequency-domain signal is defined using the Fourier-transform convention
\begin{align} \label{eq:fourier_transform}
    \tilde{x}(f) = \int_{-\infty}^{\infty} \mathrm{d}t \; x(t) \exp(-2\pi ift) .
\end{align}
The corresponding discretely sampled frequency-domain signal is denoted by
\begin{align} \label{eq:discrete_fourier_transform}
    \tilde{x}_b \equiv \tilde{x}(f_b) ,
\end{align}
where the non-negative frequencies are ${f_b\!=\! b/T}$, with ${b\!=\!0,1,\ldots,N/2}$ for even $N$.

A WDM transformation maps the time series to a set of TF-domain coefficients.
The continuous quantity
\begin{align} \label{eq:WDM_transform}
    \hat{x}(t,f)
        = \int_{-\infty}^{\infty} \mathrm{d}t' \; x(t') \, g(t';t,f) ,
\end{align}
is built from an analyzing wavepacket $g(t';t,f)$ centered at time $t$ and frequency $f$.
The discrete WDM coefficients
\begin{align} \label{eq:discrete_WD_transform}
    \hat{x}_{nm}
        \equiv
        \hat{x}(t_n,f_m) ,
\end{align}
are built from the discrete WDM basis functions
\begin{align} \label{eq:discrete_WD_basis}
    g_{nm}(t) = g(t;t_n,f_m) ,
\end{align} 
where $t_n=n\Delta T$ and $f_m \! = \! m\Delta F$.
The quantities $\Delta T \! = \! N_f\delta t$ and $\Delta F \! = \! N_t/(2T)$ are the WDM transform time and frequency resolutions respectively; these satisfy $\Delta T \Delta F \! = \! 1/2$.
The indices ${n \! = \! 0,1,\ldots,N_t-1}$ and ${m \! = \! 0,1,\ldots,N_f-1}$ label the time and frequency respectively, and ${N_t N_f \! = \! N}$.

The $N$ functions $g_{nm}(t)$ form an orthonormal basis. 
The original time series can be expanded in this basis as 
\begin{align}
    x(t) = \sum_{nm} \hat{x}_{nm} g_{nm}(t).
\end{align}

The implementation we use in Ref.~\cite{2026arXiv261117396M} requires both $N_t$ and $N_f$ to be even (hence $N$ must be doubly even).
This requirement can always be satisfied by appropriately padding the original time series.

The basis functions $g_{nm}$ are most easily defined in the frequency domain.
The definition uses $\tilde{\Phi}(f)
\! = \! \tilde{M}(f/\Delta F)/\sqrt{\Delta F}$, a scaled version of the Meyer window: 
\begin{align} \label{eq:meyer-window}
    \tilde{M}(u) =
    \begin{cases}
        1, & |u| < A, \\
        \cos\!\left[\frac{\pi}{2}\nu_d
        \!\left(\frac{|u|-A}{B}\right)\right],
        & A \leq |u| \leq A+B, \\
        0, & |u| > A+B.
    \end{cases}
\end{align}
The parameters $A$ and $B$ control the shape of the window; $A$ is the half-width of the flat-top region and $B$ is the width of the transition.
They satisfy $2A+B \! = \! 1$ (see Fig.~\ref{fig:meyer-window}). 
The normalized incomplete beta function $\nu_d(x)$ acts as a smooth transition function. 
For the full definition, see Ref.~\cite{2026arXiv261117396M}.
Unless otherwise specified, we use $A \! = \! 0.25$ (hence $B\!=\!0.5$) and $d\!=\!4$.
In the frequency domain, the WDM basis functions $\tilde{g}_{nm}(f)$ are defined as 
\begin{align} \label{eq:gnm_def_freq_domain}
    \tilde{g}_{nm}(f)
    &= \frac{1}{\sqrt{2}} \exp\!\left(-2\pi i n f \Delta T\right)
    \Bigl[
    C_{nm}\tilde{\Phi}(f + m\Delta F)
    \nonumber\\
    &\qquad\qquad
    + C_{nm}^* \tilde{\Phi}(f - m\Delta F)
    \Bigr] ,
\end{align}
where $C_{nm}\!=\!1$ if $n+m$ is even and $C_{nm}\!=\!i$ otherwise. 

\begin{figure}[t]
    \centering
    \includegraphics[width=0.49\textwidth]{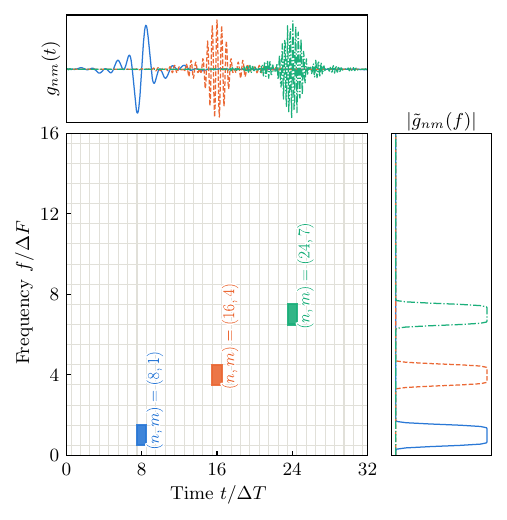}
    \caption{ \label{fig:gnm_examples}
        \emph{Top panel:} example WDM basis functions $g_{nm}(t)$ in the time domain, illustrating their localization in time.
        \emph{Right panel:} the same functions in the frequency domain, $|\tilde{g}_{nm}(f)|$, illustrating their compact support in frequency.
        \emph{Main panel:} regions of the TF plane in which the basis functions have significant support, illustrating the uniform tiling provided by the WDM transformation.
        These plots were produced for a time series of length $N=512$, with $N_f=16$ and $N_t=32$.
        This figure is based on a similar figure in Ref.~\cite{2026arXiv261117396M}, where an animated version is also available \href{https://cjm96.github.io/WDM_GW_wavelets/theory.html\#wdm-wavelets}{\faGithub}.
        } 
\end{figure}

In the time domain, this definition of the basis functions corresponds to multiplying a (co)sine wave by the inverse Fourier transform of the Meyer window $\phi(t)$. 
For example, when $n+m$ is odd,
\begin{align} \label{eq:gnm_example_odd}
    g_{nm}(t) = \sqrt{2} \sin\left(\frac{\pi m t}{\Delta T}\right) \phi(t-n\Delta T) .
\end{align}
Again, for the full definition, see Ref.~\cite{2026arXiv261117396M}. 

The case $m\!=\!0$ requires special handling in the WDM transformation.
These coefficients contain both the lowest-frequency ($f\!\sim\!0$) and highest-frequency ($f\!\sim\!f_{\rm Nyquist}\!=\!(2\delta t)^{-1}$) components of the signal, corresponding to the bottom and top rows of half-pixels in Fig.~\ref{fig:gnm_examples}.
These frequencies are often excluded from data analysis.
Nevertheless, if they are needed, the special formulae for handling these terms are discussed in Appendix~\ref{app:wdm_m0}. 

The WDM basis functions do not all have the same shape; functions with higher $m$ contain more oscillations within the window.
This reflects the uniform rectangular tiling of the TF plane provided by the WDM basis (see Fig.~\ref{fig:gnm_examples}), rather than the \emph{dyadic} tiling commonly found in other wavelet transforms.
Because conventions differ as to whether the WDM basis should be described as a wavelet basis, we avoid this terminology and instead use terms such as ``wavepacket'', ``TF pixel'', and ``atom'' throughout \cite{2020PhRvD.102l4038C, 2026arXiv260621473J}.

\subsection{Time-delay filters} \label{subsec:td_filters}

Time-delay filters, introduced in Ref.~\cite{2012JPhCS.363a2032N}, provide a method for shifting signals directly in the TF domain.

The WDM basis functions $\{g_{nm}(t)\,|\,n\in\{0,\ldots,N_t-1\},\,m\in\{0,\ldots,N_f-1\}\}$ form a complete orthonormal basis for discretely sampled time series.
Any function can be expanded using this basis; in particular,
\begin{align} \label{eq:X_expansion}
    g_{nm}(t+\delta) = \sum_{n'm'} X_{nn';mm'}(\delta) g_{n'm'}(t) .
\end{align}
The coefficients $X_{nn';mm'}(\delta)$ are the time-delay matrix elements and are functions of the constant time shift, $\delta$;
\begin{align} \label{eq:matrix_element_X}
    X_{nn';mm'} (\delta) = \int \mathrm{d} t \, g_{nm}(t+\delta)g_{n'm'}(t) .
\end{align}
These matrix elements satisfy the useful symmetry
\begin{align}
    X_{nn';mm'}(\delta) = X_{n'n;m'm}(-\delta).
\end{align}
The basis functions are localised in both time and frequency; therefore, only coefficients with $n\!\approx\! n'$ and $m\!\approx\! m'$ are significant.
The compact frequency support of the WDM basis functions means that only adjacent basis functions have nonzero overlap: the matrix elements are nonzero only for $m'\!=\!m$ or $m'\!=\!m\pm1$.

For $m,m'\!\neq\! 0$, the nonvanishing matrix elements are
\begin{align} \label{eq:X_nm}
    X_{nn';mm}(\delta) = \Re\bigg\{&(-1)^{(n-n')m}  C^*_{nm} C_{n'm} \times \\ & \exp{(2\pi i m \Delta F \delta)} T_{n-n'} (\delta)  \bigg\} , \nonumber
\end{align}
and
\begin{align} \label{eq:X_mmpm1}
    &X_{nn';m(m\pm 1)}(\delta) = \Re\bigg\{(-1)^{(n-n')m} C^*_{nm} C_{n'(m\pm 1)} \times \\ & \hspace{30pt} (\mp i)^{n-n'} \exp{\left(2\pi i \left(m \pm \frac{1}{2} \right) \Delta F \delta \right)} T'_{n-n'} (\delta) \bigg\} . \nonumber
\end{align}
The time-delay filter functions are defined as
\begin{align} \label{eq:tdfilter_T}
    T_l(\delta) = \int \mathrm{d}f \; \exp\big(2\pi i f (l \Delta T - \delta)\big) |\tilde{\Phi} (f)|^2 ,
\end{align}
and
\begin{align} \label{eq:tdfilter_Tprime}
    T'_l(\delta) = \int \mathrm{d}f & \; \exp\big(2\pi i f (l \Delta T - \delta)\big) \times \\ &\tilde{\Phi}\left(f - \frac{1}{2} \Delta F\right) \tilde{\Phi}\left(f +\frac{1}{2} \Delta F\right) . \nonumber
\end{align}
Similar expressions exist for the case when one or both of $m$ and $m'$ are zero, see the discussion in Appendix \ref{app:wdm_m0}.

\begin{figure}[t]
    \centering
    \includegraphics[width=0.49\textwidth]{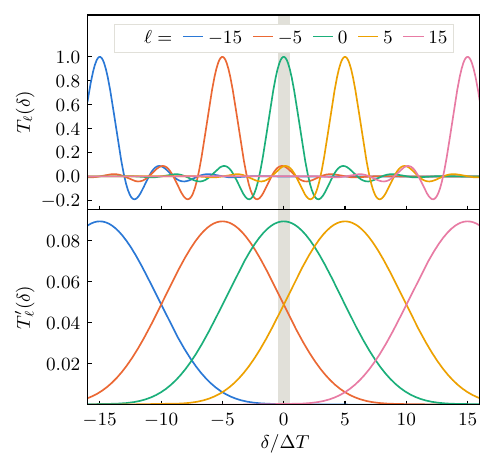}
    \caption{ \label{fig:time_delay_filters} 
        Time-delay filter functions $T_l(\delta)$ (top) and $T'_l(\delta)$ (bottom) shown for five representative values of the lag index, $l=\pm 15,\pm 5,0$.
        Because of the periodicity in Eq.~\ref{eq:delta_range}, these functions only need to be known in the range $-\Delta T/2\leq\delta<\Delta T/2$ (shaded gray region) to determine their behavior for all $\delta$.
        The plots illustrate the behavior in Eqs.~\ref{eq:delta_range} and \ref{eq:max_lag_L}.
  }
\end{figure}

The filter functions $T_l$ and $T'_l$  satisfy 
\begin{align} \label{eq:delta_range}
    T_l(\delta + \Delta T) = T_{l-1}(\delta) .
\end{align}
This means that the filters only need to be interpolated in the range $-\Delta T/2\!\leq\!\delta\!<\!\Delta T/2$; the interpolation is discussed in Appendix~\ref{app:interpolation}.
In this range the filter functions satisfy
\begin{align} \label{eq:max_lag_L}
    T_l(\delta) \rightarrow 0 \quad \mathrm{as} \quad |l| \rightarrow \infty ,
\end{align}
and similarly for $T'_l(\delta)$.
This means that the filters are only needed for $|l|\leq L$, where $L$ is the maximum lag index.
These properties can both be seen in Fig.~\ref{fig:time_delay_filters}. 
This, coupled with the dependence of the time-delay matrix elements on the lags $l \! = \! n-n'$ motivates the truncation of the sum in Eq.~\ref{eq:X_expansion} to $|l| \! \leq \! L$, which reduces the number of operations by a factor of $(2L+1)/N_t$.

It will be necessary to consider time-varying delays $\delta(t)$.
The existing formalism can be extended using
\begin{align} \label{eq:time_varying1}
    g_{nm}\left(t+\delta(t)\right) = \sum_{n'm'} Y_{nn';mm'}[\delta] g_{n'm'}(t) ,
\end{align}
where
\begin{align} \label{eq:time_varying2}
    Y_{nn';mm'}[\delta] = \int \mathrm{d} t \, g_{nm}\left(t+\delta(t)\right) g_{n'm'}(t) .
\end{align}
When the delay varies sufficiently slowly over the duration of a WDM wavepacket,
\begin{align} \label{eq:slow_vary_delay}
    \frac{1}{2\pi f_{\mathrm{eff}}\left|\dot{\delta}(t)\right|} \gg \Delta T ,
\end{align}
where $f_{\mathrm{eff}}$ is the characteristic frequency over the support of the relevant WDM basis function, we can apply the \emph{slowly-varying delay approximation}. 
Equivalently, this condition can be interpreted as the Doppler shift $f_{\rm eff}|\dot\delta|$ needing to be small compared with the frequency resolution, $\Delta F \! = \!(2\Delta T)^{-1}$.
In this approximation, the delay is treated as constant over each wavepacket and evaluated at its central time:
\begin{align}
    Y_{nn';mm'}[\delta] \approx X_{nn';mm'}\left(\delta(t_{n'}) \right) .
\end{align}

\subsection{Units} \label{subsec:units}

If the time series $x(t)$ has units $\alpha$, then the frequency-domain signal $\tilde{x}(f)$ has units $\alpha\,[\mathrm{time}]$.
The WDM coefficients $\hat{x}_{nm}$ have units $\alpha\,[\mathrm{time}]^{1/2}$, while the basis functions $g_{nm}(t)$ have units $[\mathrm{time}]^{-1/2}$.
The time-delay quantities $T_l(\delta)$, $T'_l(\delta)$ and $X_{nn';mm'}$ are dimensionless.

\section{The LISA response in the time-frequency domain} \label{sec:response}

LISA consists of 3 spacecraft (indexed by $i, j\!\in\!\{1,2,3\}$) flying in a triangular constellation.
There are 6 laser links between pairs of spacecraft. 
The optical path length (or light travel time) for the link $i\rightarrow j$ is 
\begin{align} \label{eq:light_travel_time}
    \ell_{ij}(t) = \big|\boldsymbol{x}_j(t) - \boldsymbol{x}_i\big(t-\ell_{ij}(t)\big)\big| ,
\end{align}
where $\boldsymbol{x}_i(t)$ denotes the position of spacecraft $i\in\{1,2,3\}$.
The spacecraft motions are treated nonrelativistically in a flat background spacetime.
LISA's arm lengths are $\sim\!2.5\times 10^9\,\mathrm{m}$, so the light travel times are $\sim 8\,\mathrm{s}$. 
The specific numerical orbits used in this work are described in Sec.~\ref{sec:results}.

A plane-polarized GW from a source in direction $\hat{\boldsymbol{n}}$ induces a perturbation $\delta\ell_{ij}(t)$ to the light-travel time,
\begin{align} \label{eq:time_domain_response}
    \delta \ell_{ij}(t) = \frac{1}{2} \frac{\hat{\boldsymbol{r}}_{ij}(t) \otimes \hat{\boldsymbol{r}}_{ij}(t) }{1+\hat{\boldsymbol{n}} \cdot \hat{\boldsymbol{r}}_{ij}(t) }\, : \int^{\xi_{\rm end}}_{\xi_{\rm start}} \mathrm{d}\xi \; \boldsymbol{h} (\xi) ,
\end{align}
where $\boldsymbol{h}(t)$ is the metric perturbation at
the coordinate origin (usually taken to be the solar system barycenter) and $\xi_j(t)\!=\!t+\hat{\boldsymbol{n}}\cdot\boldsymbol{x}_j(t)$
is the phase variable for the plane wave.
The integration is taken along the photon worldline between the two spacecraft, with limits $\xi_{\rm start}\!=\!\xi_i\big(t-\ell_{ij}(t)\big)$ and $\xi_{\rm end}\!=\!\xi_j(t)$.
The unit vector $\hat{\boldsymbol{r}}_{ij}(t)$ points from the sending spacecraft at the time of emission, $\boldsymbol{x}_i\big(t-\ell_{ij}(t)\big)$, to the receiving spacecraft at the time of arrival, $\boldsymbol{x}_j(t)$.
The metric perturbation is decomposed into the usual two transverse-traceless polarizations,
\begin{align} \label{eq:GW_polarizations}
    \boldsymbol{h}(t) = h^+(t) \boldsymbol{e}_+ + h^\times(t) \boldsymbol{e}_\times  .
\end{align}
For explicit forms of these see, e.g., Eq.~2 in Ref.~\cite{2003PhRvD..67b2001C}.

Equation \ref{eq:time_domain_response} is the one-arm LISA response \cite{1975GReGr...6..439E, 1985GReGr..17..719E, 1987GReGr..19.1101W, 1999ApJ...527..814A, 2003PhRvD..67b2001C, 2004PhRvD..69h2003R}.
As written, this is expressed in the time domain and is calculated from the time-domain GW waveforms, $h^{+,\times}(t)$.
The one-arm response can instead be calculated from frequency-domain waveforms $\tilde{h}^{+,\times}(f)$, using
\begin{align} \label{eq:frequency_domain_response}
    \frac{\delta \ell_{ij}(t)}{\ell_{ij}(t)} = \int^{\infty}_{-\infty} \mathrm{d}f\; \boldsymbol{D}_{ij}(2\pi f, t, \hat{\boldsymbol{n}}) : \boldsymbol{\tilde{h}}(f) e^{2 \pi i f \xi_j(t)} ,
\end{align}
where the one-arm detector tensor is defined as  
\begin{align} \label{eq:detector_tensor}
    \boldsymbol{D}_{ij}(\omega, t, \hat{\boldsymbol{n}}) = \frac{1}{2} \big(\hat{\boldsymbol{r}}_{ij}(t) \otimes \hat{\boldsymbol{r}}_{ij}(t)\big) \, \mathcal{T}_{ij}(\omega, t, \hat{\boldsymbol{n}}) ,
\end{align}
the one-arm transfer function is 
\begin{align} \label{eq:transfer_function}
    \mathcal{T}_{ij}(\omega, t, \hat{\boldsymbol{n}}) =\, & \mathrm{sinc}\left(\frac{\omega\tau_{ij}(t)}{2}\right) \exp\left(\frac{-i\omega\tau_{ij}(t)}{2}\right) ,
\end{align}
and the effective light travel time of the link is
\begin{align} \label{eq:tau_time}
    \tau_{ij}(t) = \ell_{ij}(t) \big(1 + \hat{\boldsymbol{n}} \cdot \hat{\boldsymbol{r}}_{ij} (t)\big) .
\end{align}
The quantity $\tau_{ij}(t)$ is the interval in the plane-wave phase variable $u\!=\!t+\hat{\boldsymbol{n}}\cdot\boldsymbol{x}$ that elapses during light propagation along the arm and satisfies $\tau_{ij}(t)\! \leq \! 2\ell_{ij}(t)$.

Note that $i$ and $j$ are used here to index spacecraft rather than tensor components.
Spatial tensors of any rank are denoted in boldface and, following Ref.~\cite{2003PhRvD..67b2001C}, $\otimes$ and $:$ denote the outer product and double-contraction operations, respectively:
\begin{align} \label{eq:tensor_operations}
    \boldsymbol{A}:\boldsymbol{B} = \sum_{pq} A_{pq} B_{pq} \quad\mathrm{and} \quad
    (\boldsymbol{u} \otimes \boldsymbol{v})_{pq}= u_p v_q .
\end{align}
Henceforth, to avoid an abundance of subscripts, the indices $i$ and $j$ are suppressed when this does not introduce ambiguity.
For example, we write $\delta \ell=\delta \ell_{ij}$ and $\hat{\boldsymbol{r}}=\hat{\boldsymbol{r}}_{ij}$.

The definitions of the detector tensor (Eq.~\ref{eq:detector_tensor}) and transfer function (Eq.~\ref{eq:transfer_function}) show explicitly that the LISA response depends on both time and frequency.
This contrasts with the LIGO response, which, in the usual approximation for compact binary coalescences, depends on neither time nor frequency and simply involves multiplication by an antenna pattern function.

Motivated by this, we seek to calculate the TF-domain perturbation to the arm length, $\delta\hat{\ell}_{nm}$, directly from the TF-domain GW waveforms, $\hat{h}^{+,\times}_{nm}$.
By ``directly'', we mean without first calculating the time-domain quantity $\delta \ell(t)$ and then incurring the additional cost of evaluating the discrete WDM transform to find $\delta\hat{\ell}_{nm}$.
We will exploit the fact that the TF basis functions are localized in time and have compact support in frequency to simplify the response. 

To derive the TF-domain response, start by writing the waveform as a sum of the TF basis functions and expressing all of these as functions of frequency;
\begin{align} \label{eq:metric_pol_wavelet}
    \boldsymbol{\tilde{h}}(f) = \sum_{nm} \hat{h}^+_{nm}\, \tilde{g}_{nm}(f) \boldsymbol{e_+} + \hat{h}^{\times}_{nm}\, \tilde{g}_{nm}(f) \boldsymbol{e_\times} .
\end{align}
Substituting this into the frequency-domain response in Eq.~\ref{eq:frequency_domain_response} gives 
\begin{align} \label{eq:deriv1}
    \delta \ell(t) &= \frac{1}{2} \ell(t) \sum_{nm} \big(\hat{\boldsymbol{r}}(t) \otimes \hat{\boldsymbol{r}}(t)\big) : \boldsymbol{e_+} \, \hat{h}^+_{nm} \times \\ & \quad \int^{\infty}_{-\infty} \mathrm{d}f\; \mathrm{sinc}\big(\pi f \tau(t)\big) \tilde{g}_{nm}(f) e^{2 \pi i f \left(\xi_j(t)-\tau(t)/2\right)} \nonumber \\ 
    &+\textrm{similar term for cross polarization} . \nonumber
\end{align}
The integral in Eq.~\ref{eq:deriv1} extends over all frequencies, but the basis function only has support in a band $\Delta F$ centered around $f_m$ (and, symmetrically, a band around $-f_m$).
If these bands are narrow, i.e.\ $\Delta F \! \ll \! 1/\tau(t)$, then the sinc functions can be approximated as constant across these bands. 
Because sinc is even, the constant is the same in both positive and negative bands.
In this \emph{narrowband approximation}, the response becomes
\begin{align} \label{eq:deriv2}
    \delta \ell(t) &= \frac{1}{2} \ell(t) \sum_{nm} \big(\hat{\boldsymbol{r}}(t) \otimes \hat{\boldsymbol{r}}(t)\big) : \boldsymbol{e}_+ \, \hat{h}^+_{nm} \mathrm{sinc}\big(\pi f_m \tau(t)\big) \nonumber\\ & \quad \int^{\infty}_{-\infty} \mathrm{d}f\; \tilde{g}_{nm}(f) e^{2 \pi i f \left(\xi_j(t)-\tau(t)/2\right)} \\ 
    &+\textrm{similar term for cross polarization} . \nonumber
\end{align}
The remaining frequency integral in Eq.~\ref{eq:deriv2} is a (shifted) inverse Fourier transform and evaluates to
\begin{align} \label{eq:deriv3}
    \delta \ell(t) &\approx \ell(t) \sum_{nm} F_+(t,f_m) \hat{h}^{+}_{nm} g_{nm}\big(\xi_j(t)-\tau(t)/2\big) \\
    &+\textrm{similar term for cross polarization} , \nonumber 
\end{align}
where we have defined the antenna pattern functions
\begin{align} \label{eq:new_antenna_patterns}
    F_{+}(t,f_m) = \frac{1}{2} \big(\hat{\boldsymbol{r}}(t) \otimes \hat{\boldsymbol{r}}(t)\big) : \boldsymbol{e}_{+} \, \mathrm{sinc}\big(\pi f_m \tau(t)\big) ,
\end{align}
and similarly for $F_\times(t,f_m)$.

These antenna pattern functions are time-dependent but vary slowly, on timescales of $\sim\!1\,\mathrm{year}$, due to the orbital motion of the LISA constellation.
The arm lengths $\ell_{ij}(t)$ vary on a similar timescale.
If the TF resolution $\Delta T$ is short compared to this timescale, i.e.\ $\Delta T \! \ll \! 1\,\mathrm{year}$, then the antenna pattern functions and arm lengths can be approximated as constant across each basis function; i.e.\ $F_+(t,f_m)\approx F_+(t_n,f_m)$ and $\ell(t)\!\approx \! \ell(t_n)$.
In this \emph{short-atom approximation} the response becomes
\begin{align} \label{eq:deriv4}
    \delta \ell(t) \approx \sum_{nm} \delta \hat{\ell}^{\delta}_{nm} \, g_{nm}\big(t+\delta(t)\big) ,
\end{align}
where
\begin{align} \label{eq:deltalnm_Delta}
    \delta\hat{\ell}^{\delta}_{nm} = \ell(t_{n}) \left[  F_+(t_{n}, f_m) \hat{h}^{+}_{nm} + F_\times(t_{n}, f_m) \hat{h}^{\times}_{nm} \right] .
\end{align}

Equation \ref{eq:deriv4} is almost what we need; the quantities $\delta\hat{\ell}^\delta_{nm}$ are the coefficients of the one-arm response with respect to a shifted TF basis.
The only remaining issue is that the basis functions in Eq.~\ref{eq:deriv4} are offset by a time shift that is itself dependent on time,
\begin{align} \label{eq:Delta_shift}
    \delta(t) = \boldsymbol{\hat{n}} \cdot \boldsymbol{x}_j(t) - \tau(t)/2 .
\end{align}
Using the formalism described in Sec.~\ref{subsec:td_filters} and making the \emph{slowly-varying delay approximation}, $\Delta T\ll(2\pi f|\dot{\delta}|)^{-1}$, we can write
\begin{align} \label{eq:one_arm_response_wdm}
    \delta\ell(t) = \sum_{nm} \delta \hat{\ell}_{nm} g_{nm} (t) ,
\end{align}
with 
\begin{align} \label{eq:one_arm_response_lnm}
    \delta\hat{\ell}_{nm} = \!\!\!\!\!\!\!\! \sum_{\substack{|l| \le L \\ \sigma \in \{-1,0,1\} }} \!\!\!\!\!\!\!\!  
    \delta \hat{\ell}^{\delta}_{(n-l) (m+\sigma)}  \, X_{(n-l)n;(m+\sigma)m}(\delta(t_{n})) .
\end{align}
We have used the properties of the time-delay filters to truncate the sum over the lag at a maximum index $L$.

The time shift in Eq.~\ref{eq:one_arm_response_lnm} also shifts the detector amplitude factors in Eq.~\ref{eq:deltalnm_Delta}, so they end up being evaluated at $t_n+\delta(t_n)$ instead of $t_n$, as required by Eq.~\ref{eq:deriv3}. 
This is a small effect associated with the slow time variation of these factors. 
However, to correct for it, we evaluate the detector amplitude factors at $t_n-\delta(t_n)$ prior to shifting, while retaining the original waveform coefficients.

Our implementation of the response excludes the $m\!=\!0$ coefficients which, as discussed in Appendix~\ref{app:wdm_m0}, require special handling.
As discussed above, the $m\!=\!0$ coefficients are unlikely to be used for data analysis. 
However, if they are needed, there is in principle no obstacle to calculating these terms in the response in a similar way.

A related slow-variation approximation is used when computing the gravitational-wave response of second-generation TDI observables, such as the unequal-arm Michelson combination given in Eq.~42 of Ref.~\cite{2004PhRvD..69h2001T}, for example. 
In the implementation of Eq.~20 in Ref.~\cite{2022PhRvD.106j3001K}, nested time-dependent delays are approximated by a sum of light-travel times evaluated at the same time. 
Here, we instead approximate each time-dependent shift as constant over the time resolution of a basis function.
Moreover, as the light-travel-time is dependent on the relative velocity between satellites, the error associated with the approximation is subdominant relative to the response, as $v_{\mathrm{rel}} \sim 10^{-8}$.

\subsection{Limits of applicability of the response approximations}\label{sec:response_tesselation_limits}

The response in Eq.~\ref{eq:one_arm_response_lnm} relies on three approximations: the \textit{narrowband}, \textit{short-atom}, and \textit{slowly-varying delay} approximations.
The validity of these approximations, and hence the accuracy of the response, depends on the choice of WDM time resolution and the GW frequency.
This was discussed in Sec.~\ref{sec:summary} and visualized in Fig.~\ref{fig:valid_range}.

If response accuracy were the only consideration, a smaller value of the resolution, say $\Delta T \! \sim \! 10^3\,\mathrm{s}$, would be optimal.
However, the frequency resolution, $\Delta F \! = \! (2\Delta T)^{-1}$, cannot be chosen independently and must also be sufficiently small to allow for accurate modeling of the instrumental noise.
We find that $\Delta T \! \sim \! 10^4\,\mathrm{s}$ provides a good balance between these requirements; this agrees with previous studies of the noise, e.g.\ \cite{2026arXiv261019002J} and \cite{2026PhRvD.114d4035C}, which identify similar resolutions of $\sim\!0.4\,\mathrm{days}$ and $\sim\!3200\,\mathrm{s}$ respectively as suitable. 
As shown below, this $\Delta T \! \sim \! 10^4\,\mathrm{s}$ resolution allows us to model the response sufficiently accurately across the whole LISA band.

As mentioned in Sec.~\ref{sec:summary}, the three
approximations in the zeroth-order model for the response can all be naturally extended by retaining the next term in their respective expansions.
We now describe the improvements made in this first-order model for the response.

The \textit{short-atom approximation} is improved by including terms linear in the small parameter $\Delta T/\mathrm{year}$.
Similarly, the \textit{narrowband approximation} is improved by including terms linear in $\ell/\Delta T$, or equivalently in  $2\ell\Delta F$.
These linear corrections couple a WDM atom to its neighbors through terms proportional to $t-t_n$ and $f-f_m$. 
As a result, the leading remaining errors scale quadratically with $\Delta T$ at the higher end of the $\Delta T$ range and with $\Delta F\!=\!1/(2\Delta T)$ at the lower end.

The \textit{slowly-varying delay approximation} is improved by writing $\delta(t) \!\approx\! \delta(t_{n'}) + (t-t_{n'})\dot{\delta}_{n'}$, where $\dot{\delta}_{n'}\!=\!\dot{\delta}(t_{n'})$. 
We now reexpand Eq.~\ref{eq:time_varying2} as
\begin{align} \label{eq:time_varying_first_order}
    \int \mathrm{d}t\,&g_{nm}\bigl(t+\delta(t)\bigr)g_{n'm'}(t)
    \approx \int \mathrm{d}t\,g_{nm}\bigl(t+\delta(t_{n'})\bigr)g_{n'm'}(t) \notag\\
    + \,\dot{\delta}_{n'} &\int \mathrm{d}t\, (t-t_{n'}) \, g'_{nm}\bigl(t+\delta(t_{n'})\bigr)g_{n'm'}(t) .
\end{align}
Here, $g'_{nm}$ denotes the derivative of the WDM basis function with respect to its argument.
This also introduces a new set of time-delay filters that couple neighboring WDM cells. 
The leading omitted terms are quadratic in $t-t_{n'}$ and proportional to $\dot{\delta}_{n'}^2$ and $\ddot{\delta}(t_{n'})$, respectively. 

Additionally, we note that the first term in Eq.~\ref{eq:Delta_shift} contains the motion of the center of the LISA constellation. 
We can incorporate this common R{\o}mer delay into the input waveform and apply the \textit{slowly-varying delay} approximation only to the residual delay for each link; this effectively removes the majority of the $\hat{\boldsymbol{n}}\cdot\boldsymbol{x}_j$ term from Eq.~\ref{eq:Delta_shift}.
This reduces the characteristic rate-of-change of the delay from $\sim 10^{-4}$ to $\sim 5\times10^{-7}$, reducing the variation of the delay across each WDM atom.
This final improvement can be interpreted as evaluating the waveform at the LISA constellation barycenter, rather than the solar system barycenter, and was discussed in Sec.~\ref{sec:summary}. 
As in the zeroth-order model, we evaluate the detector amplitude factors at the time each row lands after the shift, which is now $t_n$ minus the residual delay for that link, rather than $t_n-\delta(t_n)$.

\section{Results} \label{sec:results}

\begin{table}[t]
    \caption{ \label{tab:few_parameters}
        Parameters of the illustrative eccentric, prograde equatorial EMRI used to test the response.
        The source lies in the direction of M87, at the distance of the Virgo cluster; it is not
        M87's central black hole.
        The derived parameters follow from FEW's trajectory: the time to plunge is where FEW
        terminates the inspiral, and the orbit at the end of the 388-day observation record is read off it there.
    }
    \begin{ruledtabular}
        \begin{tabular}{lr}
            Parameter & Value \\
            \hline
            Primary mass, $M$ & $10^6\,\Msun$ \\
            Secondary mass, $\mu$ & $10\,\Msun$ \\
            Primary dimensionless spin, $a$ & $0.9$ \\
            Initial semi-latus rectum, $p_0$ & $8.00\,GM$ \\
            Initial eccentricity, $e_0$ & $0.3$ \\
            Initial inclination parameter, $x_0=\cos I_0$ & $1$ \\
            Luminosity distance, $D_L$ & $16.5\,\mathrm{Mpc}$ \\
            Sky colatitude, $\theta$ & $75.58^\circ$ \\
            Sky longitude, $\phi$ & $182.06^\circ$ \\
            Spin colatitude, $\theta_K$ & $60^\circ$ \\
            Spin longitude, $\phi_K$ & $60^\circ$ \\
            Initial azimuthal phase, $\Phi_{\phi,0}$ & $60^\circ$ \\
            Initial polar phase, $\Phi_{\theta,0}$ & $0^\circ$ \\
            Initial radial phase, $\Phi_{r,0}$ & $60^\circ$ \\
            FEW mode-selection threshold, $\epsilon$ & $10^{-2}$ \\
             \hline \hline
            Derived Parameters & Value \\
            \hline
            Time to plunge after the end of the observation & $0.50\,\mathrm{yr}$ \\
            Semi-latus rectum at the end of the observation & $6.02\,GM$ \\
            Eccentricity at the end of the observation  & $0.191$ \\
        \end{tabular}
    \end{ruledtabular}
\end{table}

\begin{figure}[t]
    \centering
    \includegraphics[width=0.49\textwidth]{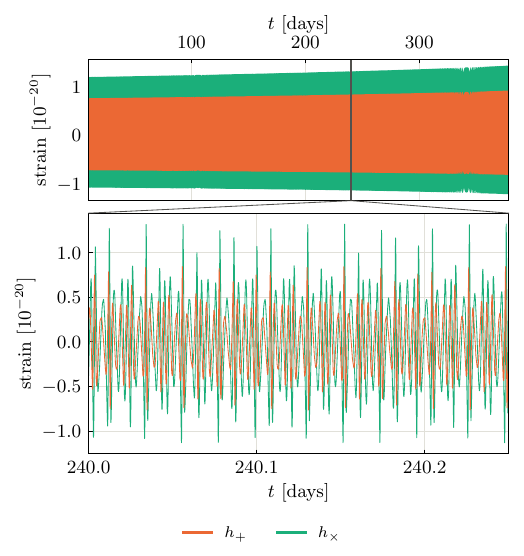}
    \caption{ \label{fig:emri-visual} 
        \emph{Top panel:} The polarizations $h_+$ (orange) and $h_\times$ (green) of the EMRI signal, over the same interval as is used in Figs.~\ref{fig:emri-response-comparison} and \ref{fig:emri-tdi-comparison} below, for consistency and ease of comparison.
        At this scale only the envelope is visible; it grows by about $20\%$ over the one-year period as the chirp progresses.
        The $h_\times$ amplitude is larger due to the viewing angle.
        \emph{Bottom panel:} A 1/4-day zoom in starting at day 240.
    }
\end{figure}

In this section, we test the TF-domain model for the LISA response by applying it to four representative LISA sources.
We compare the results with several existing implementations of the LISA response.
We also use these calculations to investigate suitable choices for the WDM resolution parameter $\Delta T$ and the maximum lag index $L$.

\begin{figure}[t]
    \centering
    \includegraphics[width=0.47\textwidth]{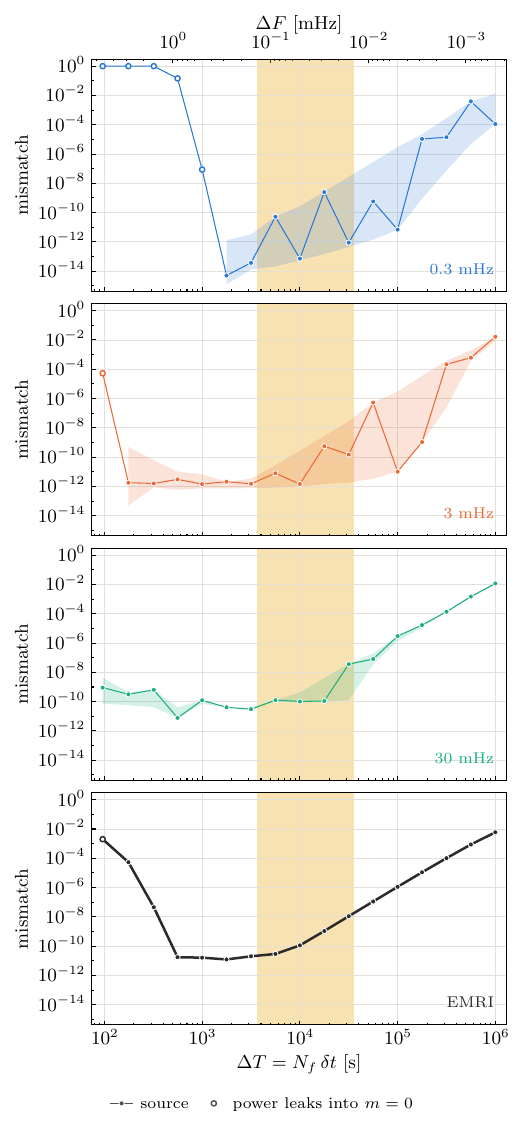}
    \caption{ \label{fig:tiling-sweep}
        Mismatches between the TF-domain response (link $1\rightarrow2$) and the time-domain reference as a function of the WDM resolution $\Delta T$.
        Results are shown for the four example sources.
        For the monochromatic signals, the shaded bands span the range of mismatches from on-bin to off-bin frequencies.  
        The yellow band shows the recommended range of resolutions from Fig.~\ref{fig:valid_range}.
    }
\end{figure}

For all numerical calculations in this section, we use the LISA orbit file,
\texttt{esa-trailing-orbits.h5}, distributed with
\texttt{lisatools} and generated using \texttt{lisaorbits}
from the ESA CReMA 1.0 Earth-trailing orbit data
\cite{2021JAnSc..68..402M}.
The spacecraft positions are expressed in heliocentric J2000 ecliptic coordinates, and the input polarizations
are prescribed at the coordinate origin.
Source sky positions are specified by the colatitude $\theta$
from the north ecliptic pole and longitude $\phi$
from the J2000 vernal equinox; these angles are fixed throughout each calculation.
For implementations that use different angular conventions,
we convert the source angles accordingly.

We first test our response using three monochromatic, circularly polarized GW signals.
These are representative of signals from galactic double white-dwarf binaries.
The GW waveforms are given by
\begin{align}
    h_+(t)&=A\cos(2\pi ft+\phi_0), \\
    h_\times(t)&=A\sin(2\pi ft+\phi_0),
\end{align}
with dimensionless strain amplitude $A=10^{-21}$ and phase offset $\phi_0=0$.
We place these sources in the direction of the Galactic center, at $(\theta,\phi)=(95.6^\circ,266.9^\circ)$.
The frequencies were chosen to be $f=3\times10^{-4}$ (low), $3\times10^{-3}$ (medium), and $3\times10^{-2}\,\mathrm{Hz}$ (high) and cover most of the LISA band.
These three sources are shown in blue, orange and green respectively in Fig.~\ref{fig:valid_range} and
in other plots in this section.

To test the response for a broadband signal with evolving frequency content, we use a fourth source: an EMRI.
The small object is on a prograde, eccentric, equatorial orbit around a spinning Kerr black hole.
We place this source in the direction of M87, at the distance of the Virgo Cluster.
The waveform was generated using version 2.0.0 of the \texttt{FastEMRIWaveforms} (FEW) package \cite{Chua:2020stf,Katz:2021yft,FastEMRIWaveformsSoftware}, with the \texttt{FastKerrEccentricEquatorialFlux} model \cite{Chapman-Bird:2025xtd,Stein:2019buj,Chua:2018woh} and the parameters listed in Table~\ref{tab:few_parameters}. 
The waveform is shown in Fig.~\ref{fig:emri-visual} and the TF harmonic tracks are shown in Fig.~\ref{fig:fig-emri-tracks}.
This source is shown in black in all figures.

For all sources, we generate time series with $N\!=\!2^{23}$ samples at a cadence of $\delta t\!=\!4\,\mathrm{s}$, corresponding to a duration of approximately $388$ days.
To suppress edge artifacts associated with the periodic extension implicit in the discrete transform, we multiply by a Tukey window with parameter $\alpha\!=\!0.05$ before zero-padding.
The zero-padding accommodates the time shifts and ensures that both $N_f$ and $N_t$ are even, as required by the WDM implementation.
Unless specified otherwise, we set the maximum lag to $L\!=\!25$ and use 128 interpolation points for the time-delay filters (see Appendix~\ref{app:interpolation}).

We first assess the accuracy of our new TF-domain model for the LISA response by comparing it with a direct time-domain implementation.
This implementation of Eq.~\ref{eq:time_domain_response} is also available in \texttt{WDM\_LISAresponse} \href{https://github.com/tomaszkinowski/WDM_LISAresponse}{\faGithub}.
The integral is evaluated using 12-point Gauss--Legendre quadrature in the retarded $\xi$ variable.
The waveforms are sampled on a uniform grid, with their values at the quadrature nodes reconstructed using centered Lagrange interpolation from the 12 nearest samples (20 for the mismatch-map tests with $f \geq 30\,\mathrm{mHz}$).
This time-domain response is not intended to be efficient; rather, it provides an accurate reference (which does not rely on the three approximations illustrated in Fig.~\ref{fig:valid_range}) against which to compare our TF-domain response.

\begin{figure*}[t]
    \centering
    \includegraphics[width=0.48\textwidth]{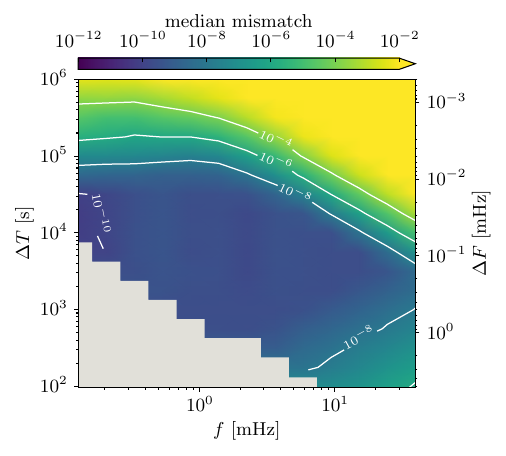}
    \hfill
    \includegraphics[width=0.48\textwidth]{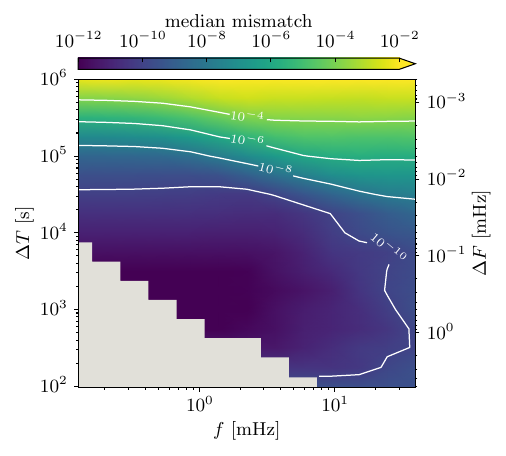}
    \caption{ \label{fig:mismatch-maps}
        Mismatch between the TF-domain response and the time-domain reference as a function of signal frequency $f$ and WDM resolution $\Delta T$.
        The secondary $y$-axis shows the corresponding WDM bandwidth $\Delta F=(2\Delta T)^{-1}$.
        Each cell shows the median mismatch for six monochromatic signals spaced evenly across the frequency band nearest $f$.
        The left and right panels show results for the zeroth- and first-order response models, respectively.
        Contours are lines of constant mismatch.
        Grey cells are excluded: at least one of their signals lies below $3\Delta F/4$, where power leaks into the $m=0$ band, which is dropped.
    }
\end{figure*}

We quantify the agreement between two time series $x$ and $y$ using the white-noise TF-domain mismatch,
\begin{align}
\mathcal{M} = 1 - \frac{|(x|y)|}{\sqrt{(x|x)(y|y)}}.
\end{align}
We do not maximize over any time or phase shifts. 
The inner product is evaluated in the WDM TF domain using
\begin{align}
    (x|y) = \sum_{nm} \hat{x}_{nm}\hat{y}_{nm}.
\end{align}
Here, $\hat{x}_{nm}$ and $\hat{y}_{nm}$ denote the WDM coefficients of the two time series. 
The sum excludes the $m\!=\!0$ frequency band and is restricted to time bins in the interior, untapered part of the Tukey window function. 
The mismatch is normalized so that it is insensitive to the overall scaling of the amplitude.
Roughly speaking, a simple indistinguishability criterion relates the accuracy needed in the response model to the signal-to-noise ratio $\rho$ (SNR) \cite{2008PhRvD..78l4020L} via the usual scaling $\mathcal{M} \! \propto \! \rho^{-2}$.
Mismatches of $\sim \! 10^{-3}$--$10^{-5}$ are generally acceptable for most GW data analysis applications, although the loudest LISA sources may require
substantially more accurate models \cite{2021PhRvD.104d4037F}.

\begin{figure}[b]
    \centering
    \includegraphics[width=0.49\textwidth]{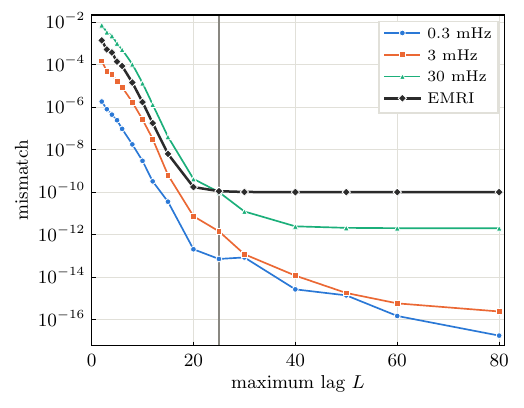}
    \caption{ \label{fig:first-lag-sweep}
        Mismatch as a function of the maximum lag $L$ for all four example sources.
        All waveforms are padded to the same length so that only the maximum lag differs between curves.
        All three monochromatic signals have on-bin frequencies (see discussion in Appendix \ref{sec:accuracy_of_zeroth_order_response}) so their mismatch decreases with $L$. 
        The EMRI signal spans many frequency bins and its mismatch plateaus at $\sim\!10^{-10}$ around $L \approx 25$.
    }
\end{figure}

All WDM responses in this section are computed using the first-order model, unless stated otherwise.
Results using the zeroth-order model are presented in Appendix~\ref{sec:accuracy_of_zeroth_order_response}.

The results for the four reference sources are shown in Fig.~\ref{fig:tiling-sweep} as a function of the WDM resolution $\Delta T$.
For the monochromatic signals, the mismatch oscillates significantly between different choices of the WDM tiling (indicated by the shaded envelope around the solid lines in the first three panels).
This behavior depends on whether the signal frequency coincides with the central frequency of a WDM atom or lies between two frequency bins.
The effect is much stronger for the zeroth-order implementation and is therefore explored in detail in Appendix~\ref{sec:accuracy_of_zeroth_order_response}.
For these comparisons, we use the one-arm response $\delta\ell_{12}$ for the $1\rightarrow2$ link.
We have verified that this link is representative of all six links.
For low frequency and small resolutions, a fraction of the signal power leaks into the $m\!=\!0$ band leading to a dramatic increase in the mismatch; points which are affected by this are indicated by open markers in Fig.~\ref{fig:tiling-sweep}.
The TF-domain response model achieves mismatches of $\mathcal{M}\!\sim\!10^{-9}$ across the LISA band.

The mismatch can be calculated for monochromatic sources as a function of $f$ and $\Delta T$.
This is shown in Fig.~\ref{fig:mismatch-maps} for both versions of the response model.
The shapes of the mismatch contours can be compared with the regions of validity in Fig.~\ref{fig:valid_range}, which were predicted from the approximations made in deriving the response.
The results in Figs.~\ref{fig:tiling-sweep} and \ref{fig:mismatch-maps} confirm that $\Delta T\!\sim\!10^{4}\,\mathrm{s}$ is a suitable TF resolution for modeling the LISA response.

Finally, we investigate how the accuracy of the TF response depends on the choice of the maximum lag $L$ used for the time shifts.
This was done for a fixed WDM resolution $\Delta T = 10^4 \,\mathrm{s}$.
The results are shown in Fig.~\ref{fig:first-lag-sweep}.
This figure shows that choosing $L=25$ is sufficient to achieve mismatches of $\mathcal{M}\!\sim\!10^{-9}$ across the LISA band.

\begin{figure}[t]
    \centering
    \includegraphics[width=0.49\textwidth]{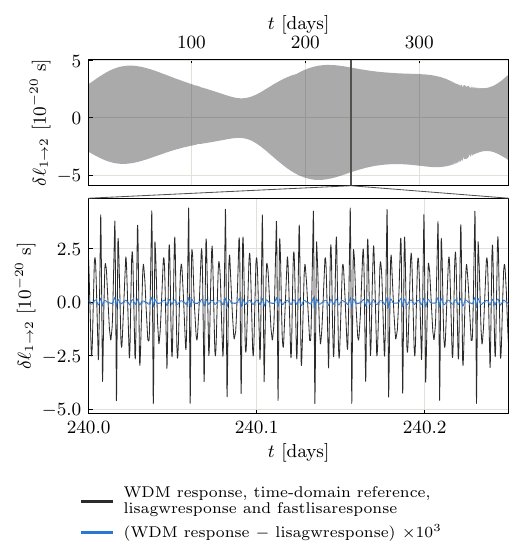} 
    \caption{ \label{fig:emri-response-comparison}
    \emph{Top panel:} The light-travel-time perturbation $\delta\ell_{1\rightarrow2}$ on link $1\rightarrow2$ for the EMRI of Table~\ref{tab:few_parameters}, over the full observation.
    Results are shown for four methods: the WDM response and time-domain reference from the \texttt{WDM\_LISAresponse} package, \texttt{lisagwresponse}, and \texttt{fastlisaresponse}.
    All four are indistinguishable at the scale shown; the largest residuals are found between the WDM TF-domain response and \texttt{lisagwresponse}.
    \emph{Bottom panel:} A quarter-day interval starting at day 240, showing the WDM response and the largest residuals, with the latter multiplied by $10^3$.
    }
\end{figure}

\subsection{Comparison with existing response codes} \label{subsec:code_comparison}

We now compare our TF-domain response model, using $\Delta T\!=\!10^4\,\mathrm{s}$, with other existing implementations of the LISA response, as well as with the time-domain reference model described above.

For these comparisons, we use \texttt{lisagwresponse} \cite{bayle_2025_17238548} and \texttt{fastlisaresponse} \cite{2022PhRvD.106j3001K}.
These packages compute the response as a fractional frequency shift which is related to the light-travel-time perturbation in Eq.~\ref{eq:time_domain_response} by a time derivative. 
Additional sign conventions and differences in the definitions of the polarization basis also need to be accounted for when making these comparisons. 

The results for the example EMRI source are shown in Fig.~\ref{fig:emri-response-comparison}.
Over the analyzed interval, the mismatch of the WDM TF-domain response model against the time-domain reference is $\mathcal{M} = 1.13\times10^{-10}$ and $\mathcal{M} = 2.26\times10^{-9}$ when compared against both external codes.
Both external codes agree with the time-domain reference to $\mathcal{M} \lesssim 3\times10^{-9}$. 
The largest residuals are found between the WDM TF-domain response and \texttt{lisagwresponse}; these residuals are shown in Fig.~\ref{fig:emri-response-comparison}.
In all cases, the residuals oscillate in phase with the input GW signal; there is no sign of any secular errors that accumulate over the duration of the observation.

We note that both external packages use a different formula for the response that omits additional subdominant terms due to the spacecraft velocities. 
In particular, differentiating Eq.~\ref{eq:time_domain_response} introduces endpoint factors involving $1+\hat{\boldsymbol n}\cdot\boldsymbol v_{i/j}$.  
We have checked that by making additional changes to the external-code outputs designed to approximately account for these effects, we further lower the mismatch against the time-domain reference to $\mathcal{M} \approx 9.7 \times 10^{-11}$ for both implementations. 
However, we chose to not pursue these corrections any further here as mismatches of $\mathcal{M}\!\sim\!10^{-9}$ (for example, as reported in Fig.~\ref{fig:tiling-sweep}) are already expected to be low enough.

\begin{figure}[t]
    \centering
    \includegraphics[width=0.49\textwidth]{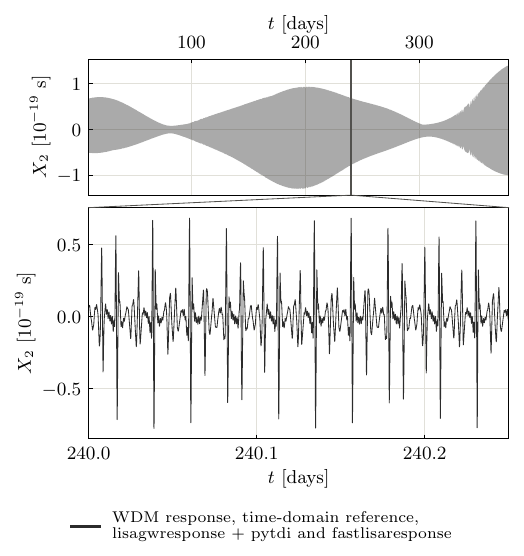}
    \caption{ \label{fig:emri-tdi-comparison} 
        \emph{Top panel:} The second-generation Michelson channel $X_2$ for the EMRI signal over the full observation, computed using four methods: the WDM response and time-domain reference from the \texttt{WDM\_LISAresponse} package, \texttt{lisagwresponse} (with $X_2$ calculated using \texttt{pytdi}), and \texttt{fastlisaresponse}.
        All four results are indistinguishable at the scale shown; only the WDM response is plotted, shown here as its envelope.
        \emph{Bottom panel:} A quarter-day interval starting at day 240, again showing only the WDM result because the four curves are indistinguishable at this scale.
    }
\end{figure}

\subsection{Construction of TDI variables in the TF domain}

The LISA one-arm measurements are dominated by laser frequency noise, necessitating the use of TDI variables for data analysis.
The TDI variables are constructed from the one-arm responses by multiple time-shift operations which can be carried out in the TF domain using the methods described in Sec.~\ref{subsec:td_filters}.
For this reason, we have chosen to focus on showing results for the one-arm response throughout most of this paper.
Nevertheless, we show a single proof-of-concept example of a TDI response calculated in the TF domain in this section.

Specifically, we use the second-generation Michelson-like TDI variables $X_2$, $Y_2$, and $Z_2$, which account for the time dependence of the constellation's arm
lengths \cite{2003PhRvD..68f1303S,2004PhRvD..69h2001T}.
These are calculated for the example EMRI source using the TF-domain model for the response with $\Delta T\!=\!10^4\,\mathrm{s}$.
The results are shown in Fig.~\ref{fig:emri-tdi-comparison} for all four of the response models described in Sec.~\ref{subsec:code_comparison}. 
In all cases, the responses are indistinguishable at this scale and have mismatches comparable to those for the one-arm responses described in Sec.~\ref{subsec:code_comparison}.

\section{Discussion} \label{sec:discussion}

This paper has presented an approach to computing the LISA response to GWs directly in the time-frequency domain using the WDM basis.
This approach enables several natural approximations that simplify and accelerate the calculation of the response.
With the first-order approximation described here, the response can be calculated with high accuracy for signals across the LISA band, with mismatches of $\mathcal{M}\!\sim\!10^{-9}$.
The zeroth-order version of the response has been fully implemented in the public \texttt{JAX}-based Python package \texttt{WDM\_LISAresponse} \href{https://github.com/tomaszkinowski/WDM_LISAresponse}{\faGithub}, while the improved first-order version is under active development.

The accuracy of the response depends on the choice of WDM time resolution.
We have investigated this dependence for a range of representative sources and find that $\Delta T\!\sim\!10^{4}\,\mathrm{s}$ provides a suitable resolution for accurately modeling the response.
Conveniently, this resolution has also previously been shown to be suitable for modeling the LISA noise.

We have also demonstrated a proof-of-concept implementation of TDI variables constructed directly in the TF domain.
Further work is ongoing to efficiently implement this TF-domain TDI construction in a \texttt{JAX}-based package suitable for use in a likelihood function for a global fit.

An important consideration for the use of the response for data analysis (e.g.\ as part of a global fit) is the computational cost. 
As the \texttt{WDM\_LISAresponse} package has been written in \texttt{JAX}, it supports just-in-time compilation, 
which translates to overhead for the first evaluation, with subsequent ones reusing the compiled code. 
For $L = 25$ and $\Delta T = 10^4\,\mathrm{s}$ and a signal of duration $T = 388\,\mathrm{days}$, the first evaluation of the zeroth-order model, including compilation, from WDM waveform input to six-link output, takes $3.56 \,\mathrm{s}$, with subsequent evaluations taking $\sim 0.539\,\mathrm{s}$. 
These timings assume that $\Delta T$ and $L$ are fixed between runs.
These measurements were performed using an Intel Core Ultra 7 265K CPU with 20 logical cores. 
For comparison, \texttt{fastlisaresponse} and \texttt{lisagwresponse} achieve runtimes of approximately $5.11\,\mathrm{s}$ and $27.9 \,\mathrm{s}$, respectively, from waveform input to six-link output. 
However, \texttt{fastlisaresponse} is a GPU-optimized code – on an NVIDIA RTX A400 GPU the same calculation takes $2.84 \,\mathrm{s}$.

Further modifications to the \texttt{WDM\_LISAresponse} package may improve its performance on GPUs and are currently under investigation.
Preliminary tests of these modifications indicate that, for the same WDM parameters and signal duration, subsequent evaluations can achieve runtimes of $0.0567\,\mathrm{s}$, while the first evaluation, including compilation, takes $3.66\,\mathrm{s}$.
These measurements were performed on an NVIDIA RTX A400 GPU.

Further runtime improvements may be possible by exploiting the localization of signal power for monochromatic and multiharmonic waveforms along specific tracks in the TF grid.
A sparse implementation of the response model that accounts for the overlap between these tracks could reduce the computational cost of evaluating multiple waveforms in parallel.
We plan to investigate how best to exploit this sparsity to improve runtime while preserving accuracy.

The TF-domain response should be viewed as one component of a larger pipeline, illustrated in Fig.~\ref{fig:flow_chart}, for performing LISA data analysis consistently in the TF domain from end to end.
The results presented here demonstrate that the LISA response can be modeled accurately and efficiently in the TF domain, establishing the viability of an important component of this framework.


\section*{Data Availability}

The WDM transformations in this work have been computed using the package \texttt{WDM\_GW\_wavelets} \href{https://github.com/cjm96/WDM_GW_wavelets}{\faGithub}~\cite{2026arXiv261117396M,WDM_github}. 
All figures in the paper have been generated using the \texttt{paper\_figures} branch of \texttt{WDM\_LISAresponse} \href{https://github.com/tomaszkinowski/WDM_LISAresponse}{\faGithub}.
In building and testing the \texttt{WDM\_LISAresponse} package, we have used the Python packages \texttt{NumPy}~\cite{2020Natur.585..357H}, \texttt{SciPy}~\cite{2020NaMet..17..261V}, \texttt{JAX}~\cite{jax2018github}, \texttt{Matplotlib}~\cite{2007CSE.....9...90H}, \texttt{h5py}~\cite{h5py}, \texttt{LISA Constants}~\cite{LISAConstants}, \texttt{LISA Orbits}~\cite{LISAOrbits}, \texttt{LISA Analysis Tools}~\cite{LISAAnalysisTools}
and \texttt{FastEMRIWaveforms}~\cite{Chua:2020stf,Katz:2021yft,FastEMRIWaveformsSoftware}.
For validation against existing response codes, we have used \texttt{lisagwresponse}~\cite{bayle_2025_17238548}, \texttt{PyTDI}~\cite{Staab2026PyTDI} and \texttt{fastlisaresponse}~\cite{2022PhRvD.106j3001K}.

\begin{acknowledgments}
    We thank all the members of the UK LISA ground segment (GS) team for very useful ongoing discussions about this work.
    We also thank James Alvey for comments on a draft version of the manuscript.
    TK acknowledges support from the Science and Technology Facilities Council through a Cambridge Astronomy DTP 2025 studentship [1732].
    CJM acknowledges support from the UKSA supporting the UK’s contribution to LISA Ground Segment activities.
    We acknowledge the LISA Simulation Expert Group for providing the LISA simulation software used in this work.
    We thank the Institute of Astronomy, University of Cambridge, for providing the computing resources used for the numerical comparisons and benchmarks in this work.
    Parts of the software package and the code for making figures were developed with the assistance of Claude Opus 5.5, and ChatGPT-6.1 Sol was used to help with drafting parts of the text; in all cases, the output was carefully reviewed by the human authors.
\end{acknowledgments}


\bibliographystyle{apsrev4-1}
\bibliography{references}


\appendix

\section{m = 0 WDM coefficients} \label{app:wdm_m0}

The definition of the WDM $g_{nm}$ basis functions given in Eq.~\ref{eq:gnm_def_freq_domain}, or equivalently in Eq.~\ref{eq:gnm_example_odd}, only applies for $0<m<N_f$. 
This leaves the lowest frequencies around $f \! \sim \! 0$ and the highest frequencies around $f \! \sim \! f_{\rm Nyquist}$ that have to be handled differently. 
In the WDM transformation, these endpoint basis functions have the alternative forms 
\begin{align}
    &g_{n0} = \phi\left(t-2n\Delta T\right ) ,  \nonumber \\ 
    & g_{nN_f} = \cos{\left(2 \pi f_{\mathrm{Nyquist}} (t- 2n \Delta T) \right)} \phi(t- 2n \Delta T) .
\end{align}
Unlike the interior basis functions, they are shifted by twice the resolution, $2\Delta T$, and there are therefore only $N_t/2$ independent values of the $n$ index.
This can be conveniently handled by combining both the low and high frequency terms into a single $m=0$ row of the TF transform using the first half for the zero-frequency terms and the second half for the Nyquist-frequency terms.
This preserves the total number $N$ of WDM coefficients, while keeping a uniform $N_t\times N_f$ coefficient grid. 
Graphically, this can be seen represented in Fig.~\ref{fig:gnm_examples} by the half-width pixels along the top and bottom rows. 
The details of this implementation are described in Refs.~\cite{2026arXiv261117396M, WDM_github}.

As the boundary coefficients at $m=0$ are defined differently, the corresponding time-shift operators have to be treated separately. Hence, they are omitted from the current implementation. 
Consequently, the implementation of the LISA response described in this paper accurately models waveform spectral features in the range:
\begin{align}
    3\Delta F/4 \lesssim f \lesssim f_{\mathrm{Nyquist}}-3\Delta F/4.
\end{align}

\section{Time-Shift Implementation in \texttt{WDM\_GW\_wavelets}} \label{app:interpolation}

Following the description in Sec.~\ref{subsec:td_filters}, we calculate the time-delay filter functions $T_l(\delta)$ and $T'_l(\delta)$ for the $2L+1$ lag values $-L\leq l\leq +L$ over a uniform grid of $P=128$ points in $u =\delta/\Delta T \in [-1/2, 1/2]$.
Since the filters are a function of the dimensionless variable $l-\delta/\Delta T$, they are computed independently of tiling choice $N_t, N_f$. 
The integrals are evaluated via Gauss-Legendre quadrature on each interval where the window is smooth, with agreement to $\sim 10^{-12}$ with adaptive quadrature.

During the time shift operation, the filters are evaluated at each row's delay $\delta_n$ by linear interpolation.
Delays outside $\pm\Delta T/2$ are folded back by shifting the lag index.
Linear interpolation on a uniform grid is straightforward to compile and differentiate in \texttt{JAX}. Its error scales as $(P-1)^{-2}$.
For $P = 128$ the filters are reproduced to within $\sim 7\times 10^{-6}$ of their peak values, and the estimated mismatch contribution in the off-bin tone test is $\lesssim 10^{-10}$. 
Therefore, this choice of $P$ is conservative.
Note that $P$ sets the initial cost of constructing the  data to be interpolated and its memory footprint; it does not impact the per-waveform evaluation cost.

The interpolated filters build the time-delay matrices of Eq.~\ref{eq:one_arm_response_lnm}, applied as a sum over the lag $l$.
At each $l$, all three bands $\sigma\in\{-1,0,1\}$ are assembled together over the full $\left(N_t,N_f\right)$ grid.
Hence the total number of operations scales approximately as $3N(2L+1)$.

\section{Accuracy of the zeroth-order response model}\label{sec:accuracy_of_zeroth_order_response}

\begin{figure}[t]
    \centering
    \includegraphics[width=0.48\textwidth]{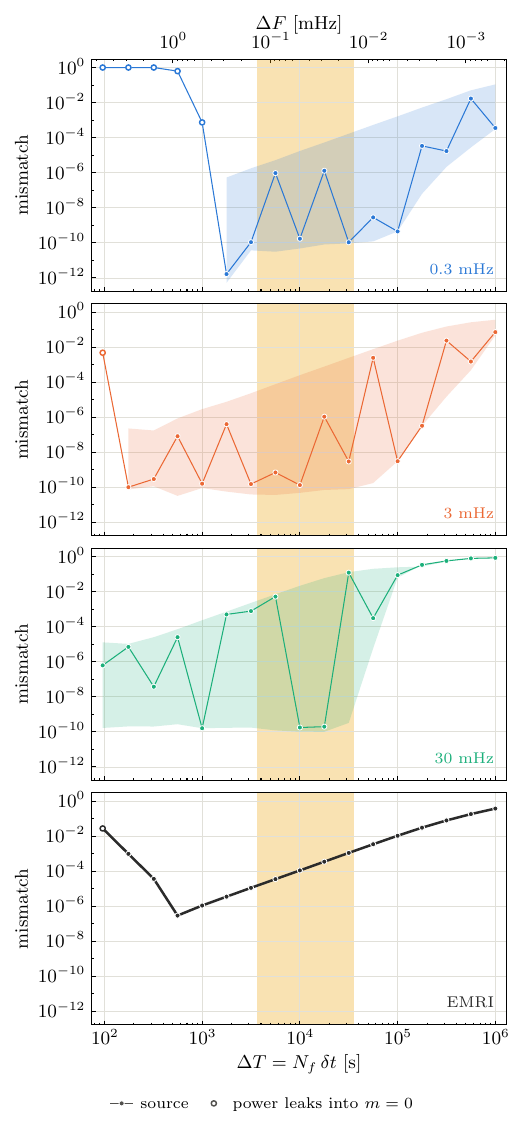}
    \caption{ \label{fig:tiling-sweep-zeroth} 
    Mismatches between the TF-domain zeroth-order response model
    (link $1\rightarrow2$) and the time-domain reference, as a function of the WDM time resolution
    $\Delta T = N_f\,\delta t$.
    The setup details of the figure are the same as in Fig.~\ref{fig:tiling-sweep}.
    }
\end{figure}

In this section we investigate the accuracy of the zeroth-order response model, shown in Fig.~\ref{fig:tiling-sweep-zeroth}, with the same setup as in Sec.~\ref{sec:results}.

\begin{figure}[t]
    \centering
    \includegraphics[width=0.49\textwidth]{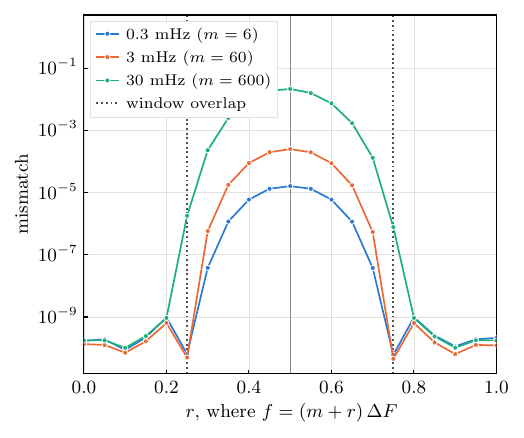}
    \caption{ \label{fig:off-bin} Mismatch as each white-dwarf binary is moved across
    a frequency band, $f = (m + r)\,\Delta F$, at $\Delta T = 10^4\,\mathrm{s}$
    ($N_f = 2500$), labeled by its frequency at $r = 0$; everything else is as in
    Fig.~\ref{fig:tiling-sweep-zeroth}. For $r < A$ or $r > 1 - A$ (dotted lines) the signal
    lies on the flat top of a single band's Meyer window; between them it lies in the
    window overlap of bands $m$ and $m + 1$. The mismatch stays at $10^{-10}$--$10^{-9}$
    on the flat top and peaks off-bin, at $r = 1/2$ (solid line).}
\end{figure}

\begin{figure}[b]
    \centering
    \includegraphics[width=0.49\textwidth]{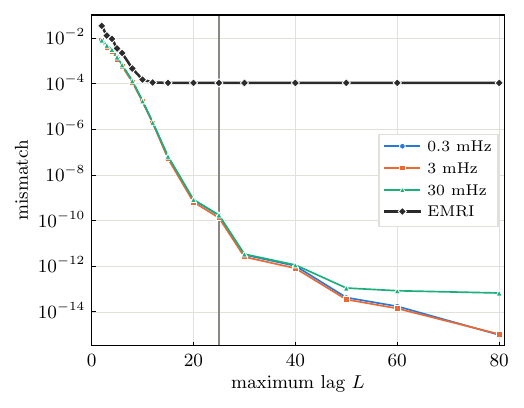}
    \caption{ \label{fig:zeroth-lag-sweep}
        Mismatch against the maximum lag $L$ for all four reference sources.
        All waveforms are padded to the same length so that only the maximum lag differs between curves.
        All three monochromatic signals have an on-bin frequency and their mismatch falls steeply with $L$, from $\sim 10^{-2}$ at $L = 2$ to $\sim 2\times10^{-10}$ at $L = 25$ (vertical line, used throughout) and $\sim 10^{-15}$ at $L = 80$ ($\sim 7\times10^{-14}$ for the $30\,\mathrm{mHz}$ signal). 
        The EMRI stops improving at $L \approx 15$, at $10^{-4}$, where its off-bin content dominates.
    }
\end{figure}

For monochromatic signals, we observe that the mismatch values strongly oscillate between different tiling choices (solid lines in the first three panels), as a fixed frequency lands at a different offset $r$ within its band at each tiling. 
As in the first-order response model, this is attributed to whether the frequency of a monochromatic signal lies on the centre of a WDM band (on-bin), or in the region between two band centres (off-bin), which is illustrated in Fig.~\ref{fig:off-bin} for a single tiling, $\Delta T = 10^4\,\mathrm{s}$. 
As the frequency moves away from an integer multiple of $\Delta F$, the mismatch increases by five to eight orders of magnitude for off-bin signals.

The discrepancy between on-bin and off-bin signals is attributed to the shape of the Meyer window (see Eq.~\ref{eq:meyer-window} and Fig.~\ref{fig:meyer-window}) and the \textit{slowly-varying delay approximation}. 
The error from neglecting the delay variation can be interpreted as a frequency shift $f \dot\delta_n$, which changes the window weight by a term proportional to $f\dot\delta_n\tilde{\Phi}'$. 
On the flat top ($r < A$ or $r > 1-A$), $\tilde{\Phi}' = 0$, so this leading-order error is suppressed for on-bin signals; in the window overlap $\tilde{\Phi}' \neq 0$, so off-bin signals are affected more significantly. 
This is further corroborated by the fact that in Fig.~\ref{fig:off-bin}, the mismatch grows when the frequency enters the transition regions (Eq.~\ref{eq:meyer-window} and Fig.~\ref{fig:meyer-window}).

We observe that the mismatch at the top end of the shaded regions scales approximately as $\Delta T^2$ at fixed $f$, and grows with $f$: the frequency shift $f\dot\delta_n$ is measured against the band width $\Delta F = 1/(2\Delta T)$, so the error grows as $f\Delta T$, in line with the \textit{slowly-varying delay approximation}. 
The lower end remains at a floor set by the maximum lag $L$ (see Fig.~\ref{fig:zeroth-lag-sweep}), and increases at coarser $\Delta T$. 
This can be attributed to the fact that at coarse tilings, the delay-induced frequency shift becomes comparable to the width of the flat-top region as $\Delta F$ decreases, $f|\dot\delta|_{\max} \gtrsim A\,\Delta F$. 
With $|\dot\delta|_{\max} \approx v/c \approx 10^{-4}$, this happens at $\Delta T \approx 4\times10^{4}\,\mathrm{s}$ for the $30\,\mathrm{mHz}$ signal, where its lower end jumps by four orders of magnitude.

As in Fig.~\ref{fig:tiling-sweep}, the EMRI's evolving frequency over the analyzed duration sweeps between on-bin and off-bin values, explaining why the mismatches do not reach the floor achieved by monochromatic signals. 
Between $\Delta T \!\approx\! 5.6\times10^{2}\,\mathrm{s}$ and $1.8\times10^{5}\,\mathrm{s}$, the EMRI mismatch stays at $0.35$--$0.48$ of the off-bin bound of the $3\,\mathrm{mHz}$ signal, close to its dominant power ($f=2$--$5\,\mathrm{mHz}$), and therefore follows the same $\Delta T^2$ scaling, with $\mathcal{M} \!=\! 1.1\times10^{-4}$ at $\Delta T = 10^4\,\mathrm{s}$. 
Below $\Delta T \approx 5.6\times10^{2}\,\mathrm{s}$, the mismatch rises again as the EMRI's power reaches the excluded $m=0$ band.

Investigations into the variability of accuracy of the zeroth-order response model with maximum lag index (for $\Delta T = 10^4 \,\mathrm{s}$) in Fig.~\ref{fig:zeroth-lag-sweep} indicate that the lower bounds in Fig.~\ref{fig:tiling-sweep-zeroth} are limited by the choice of parameter $L$: at $L = 25$, the on-bin mismatches of $1.3$--$1.8\times10^{-10}$ in Fig.~\ref{fig:zeroth-lag-sweep} match the lower bounds at this tiling. 
As the EMRI signal contains power at both on-bin and off-bin frequencies, the mismatch is dominated by off-bin contributions, which are insensitive to increasing the maximum lag parameter: the EMRI mismatch stays at $1.1\times10^{-4}$ from $L \approx 15$ onwards, the value of the $\Delta T^2$ trend in Fig.~\ref{fig:tiling-sweep-zeroth} at this tiling.

\end{document}